\documentclass[12pt]{article}

\usepackage[top=50mm, bottom=50mm, left=50mm, right=50mm]{geometry}
\usepackage{lineno}
\usepackage{comment}
\usepackage{amssymb}
\usepackage{amsmath}
\usepackage{amsthm}
\usepackage{epsfig}
\usepackage{graphicx}
\usepackage{graphics}
\usepackage{float}
\usepackage{subfigure}
\usepackage{multirow}
\usepackage{color}
\usepackage{lineno}
\usepackage{fullpage}
\usepackage[normalem]{ulem}
\usepackage{makeidx}
\usepackage{xspace}
\usepackage{wrapfig}
\makeindex

\definecolor{darkred}{rgb}{1, 0.1, 0.3}
\definecolor{darkblue}{rgb}{0.1, 0.1, 1}
\definecolor{darkgreen}{rgb}{0,0.6,0.5}

\newcommand{\E}[1]		{{ \textcolor{red} {{\sc Elena Says:} #1}}}
\newcommand{\Ho}[1]		{{ \textcolor{blue} {{\sc Hoang Says:} #1}}}
\newcommand{\Ol}[1]		{{ \textcolor{green} {{\sc Olha Says:} #1}}}

\newtheorem{definition}{Definition}

\allowdisplaybreaks[1]
\usepackage[flushleft]{threeparttable}

\definecolor{dkgreen}{rgb}{0,0.4,0}
\definecolor{gray}{rgb}{0.5,0.5,0.5}
\definecolor{mauve}{rgb}{0.58,0,0.82}

\usepackage{scalerel,stackengine}
\stackMath
\newcommand\reallywidehat[1]{%
\savestack{\tmpbox}{\stretchto{%
  \scaleto{%
    \scalerel*[\widthof{\ensuremath{#1}}]{\kern-.6pt\bigwedge\kern-.6pt}%
    {\rule[-\textheight/2]{1ex}{\textheight}}
  }{\textheight}%
}{0.5ex}}%
\stackon[1pt]{#1}{\tmpbox}%
}
\begin{document}

\title{Monitoring the Dynamic Networks of Stock Returns}
\author{Elena Farahbakhsh Touli$^{(a)}$, Hoang Nguyen$^{(b)}$, Olha Bodnar$^{(b)}$\\
$^{(a)}$ Department of Mathematics - Stockholm University, Sweden\\
$^{(b)}$ School of Business - Örebro University, Sweden }

\date{\today}
\maketitle

\begin{abstract}
In this paper, we study the connection between the companies in the Swedish capital market. We consider 28 companies included in the determination of the market index OMX30. The network structure of the market is constructed using different methods to determine the distance between the companies. We use hierarchical clustering methods to find the relation among the companies in each window. Next, we obtain one-dimensional time series of the distances between the clustering trees that reflect the changes in the relationship between the companies in the market over time. The method from statistical process control, namely the Shewhart control chart, is applied to those time series to detect abnormal changes in the financial market.
 \\

\end{abstract}

\newpage

\section{Introduction}
Financial market is often considered as a network where the nodes are companies and the links among nodes represent the connectedness \cite{Diebold2015}.
The connectedness of financial assets plays an important role for policy makers and forecasters, especially during recessions and crises, see \cite{Minoiu2015}, \cite{Bouri2021}, among others. However, there is a controversy on how to define and measure the connectedness as well as how to keep track of the changes in it. \cite{Vandewalle2001}, \cite{Onnela2004}, \cite{Bonanno2004} and \cite{Chi2010} use the Pearson correlation of financial returns as a measure of connectedness in the network. This measure is symmetric and may be subject to the choice of the sample size. Recently, \cite{Diebold2014} introduce an asymmetric measure of connectedness based on the effects of a shock from one node to other nodes.

In this study, we take into account the two common measures of the connectedness in the networks, namely the Pearson Correlation Coefficient Dissimilarity (PCCD) and the Generalized
Variance Decomposition Dissimilarity (GVDD). First, we contribute to the literature by comparing the center of the network determined by these measures. Second, we use a hierarchical clustering method to divide the dense networks into sparse trees. Using the tree representation of the network, it is easier to analyse which companies are more related to each others. Third, we monitor the real time changes in the tree distance as a signal of changes in the financial network. Lastly, we analyze 28 biggest companies listed in Sweden stock exchange to illustrate the pros and cons of the considered connectedness measures.

The center of the network is one of the major interests as it is a node or group of nodes that makes most influential to other nodes in the network.
\cite{Diebold2014} analyzes the volatility connectedness and define the centrality in term of the net transmitter of shock. On the other hand, we model the financial returns as a network graph where the distance in the graph is based on the connectedness/dissimilarity measures. Hence, the center of the network can be considered as the node with a shortest distance to the furthest node. We also analyze how the center changes and evolves over time.

Another interesting feature of the financial network is to identify which neighbours or group of companies that are most related to each other. Similar to \cite{Mantegna1999} we use a hierarchical clustering method to convert the networks of companies to rooted trees. The hierarchical clustering algorithm takes advantage of the distance in the network to merge the most similarity nodes into a cluster. The tree structure also highlights the main difference of the implied network based on the two connectedness measures.

In \cite{Diebold2014}, the changes in the network are monitored by using a total sum of connectedness. But that measure ignores the entity of the total changes and somehow misinterprets the real changes in the network structure. To get over this shortcoming, we use a tree distance method \cite{ITGRFMfCPT} based on the generalized Robinson-Foulds distance \cite{Nye2006}. The generalized Robinson-Foulds compares two trees and pairing splits in one tree with similar splits in the other. Hence, we obtain a daily series of distance due to the changes in the tree structures. We spot out several abnormal jumps of changes and recommend it as a warning signal.

Using the returns on stocks traded on the Swedish capital market, we analyze the network of the financial returns during the last five years from 2017 to 2022. We consider a rolling windows of three months to calculate the daily measures of connectedness. It appear that {\it Investor}, a Swedish investment company, is the center of the network in most of time for both considered connectedness measures. However, there is a quite difference of hierarchical clustering trees between two measures. In general, the companies in the same sector are closer together, but the links between sectors diverse between these two methods. We also observe that the tree distance computed using \cite{Diebold2014} are in a higher magnitude and more volatile than the one by Pearson correlation. However, the tree distance are correlated and inline with the high volatility period of stock returns.

The rest of the paper is organized as follows. Section 2 describes the two common measures of connectedness or dissimilarity. Section 3 introduces the network construction of financial returns by applying the considered measures. Here, we outline the notations of a center in a network as well as describe how to form a hierarchical clustering tree and to compute tree distance. An empirical illustration is presented in Section 4 and conclusions are reached in Section 5.


\section{Dissimilarity measure of stock returns} \label{Sec:dissimilarity}

In this section, we present two common connectedness measures of financial returns based on the Pearson correlation and variance decomposition.

Let $p_{i,t}$ be the closing price of stock $i$ at day $t$ and let denote $r_{i,t} = \log(p_{i,t}) - \log(p_{i,t-1})$ denote the log-return on stock $i$ on day $t$. The Pearson correlation coefﬁcients (PCC) between asset returns $i$ and $j$ at time $t$ is defined using data over $M$ periods from $t_0 = t-M+1$ upto $t$ as
\begin{equation}
\begin{aligned}
\rho_{i,j}^{t} = \frac{ \displaystyle\sum_{s = t_0}^t \left( r_{i,s} - \bar{r}_i \right) \left( r_{j,s} - \bar{r}_{j} \right)  }{ \sqrt{\displaystyle\sum_{s = t_0}^t \left( r_{i,s} - \bar{r}_i \right)^2 \displaystyle\sum_{s = t_0}^t \left( r_{j,s} - \bar{r}_{j} \right)^2 }},
\end{aligned}
\end{equation}
where $\bar{r}_i=\dfrac{1}{M}\displaystyle\sum_{s = t_0}^t r_{i,s}$ and $\bar{r}_j=\dfrac{1}{M}\displaystyle\sum_{s = t_0}^t r_{j,s}$ are the sample means of the returns on the $i$ and $j$ stocks, respectively, computed over the last $M$ periods. The correlation coefficient receives a value in the range $[-1,1]$ and it represents the linear dependence between two financial returns. Then, the dissimilarity $h_{i,j}^{t,PCCD}$ between two stock returns $i$ and $j$ at time $t$ can be written as
\begin{equation}\label{Eq:distanceMatrix}
\begin{aligned}
h_{i,j}^{t,PCCD} = \sqrt{2 (1 - \rho_{i,j}^{t})}.
\end{aligned}
\end{equation}
This definition of the dissimilarity satisfies the three axioms that define a metric $\mathbf H^{t, PCCD}=[h_{ij}^{t, PCCD}]$ (see, \cite{Mantegna1999}) where $0 \leq h_{i,j}^{t,PCCD} \leq 2$. The dissimilarity expresses the level at which the stocks are correlated (e.g., \cite{Onnela2003}).

The Pearson Correlation Coefficient Dissimilarity (PCCD) in \eqref{Eq:distanceMatrix} measures a relationship among variables and has been employed in many studies, for example \cite{Vandewalle2001}, \cite{Onnela2004}, \cite{Bonanno2004} \cite{Chi2010}, among others. The PCCD only considers the pairwise linear correlation but ignores other nonlinearities created by the time-varying correlations. It is also a nondirectional measure that makes it difficult to distinguish the asymmetric effect of one firm to another. \cite{Diebold2014} propose a measure of the similarity based on the variance decomposition associated with a VAR model which helps to overcome the limitation of the PCCD. The similarity matrix is created based on the shares of forecast error of the returns which allows to measure how many percentages of the forecast error of one variable caused by another variable.

Following \cite{Diebold2014}, a VAR model of order $p$ is used to model the dynamic behaviour of asset returns expressed as
\begin{equation}\label{VAR}
\begin{aligned}
\mathbf r_{t} & = \mathbf B_1 \mathbf r_{t-1} + \ldots + \mathbf B_p \mathbf r_{t-p} + \mathbf \Sigma^{1/2} \boldsymbol \epsilon_t,
\end{aligned}
\end{equation}
where $\mathbf r_{t}$ is a $n$-dimensional vector of demean asset returns; $\mathbf B_j$ is a $n \times n$ variate matrix of regression coefficients with $j = 1, \ldots, p$;
$\mathbf \Sigma$ is a  $n \times n$ covariance matrix that describes the interaction between the components of the error process;
$\boldsymbol {\epsilon}_t$ is a $n$-dimensional vector of error terms that follows a white noise process with zero mean vector and identity covariance matrix.

We rewrite the VAR model \eqref{VAR} in the moving average (MA) representation as
\begin{equation}\label{MA}
\begin{aligned}
\mathbf r_{t} & = \boldsymbol \Theta( \boldsymbol L) \boldsymbol \Sigma^{1/2} \boldsymbol \epsilon_t , \\
\boldsymbol \Theta( \boldsymbol L) &= \left( \boldsymbol I - \boldsymbol B_1 \boldsymbol L - \ldots \boldsymbol B_p \boldsymbol L^p \right)^{-1} = \boldsymbol \Theta_0 +  \boldsymbol \Theta_1  \boldsymbol L + \boldsymbol \Theta_2  \boldsymbol L^2 + \ldots ,
\end{aligned}
\end{equation}
where $L$ is the lag operator that is $L \mathbf r_t=\mathbf  r_{t-1}$. To compute the MA representation, the Cholesky factor of the covariance matrix $\boldsymbol \Sigma$ is commonly used together with the generalized variance decomposition (GVD) framework of \cite{Koop1996} and \cite{Pesaran1998}. Note that the GVD helps to produce variance decompositions that is invariant to the order of the variables. \cite{Diebold2014} consider the standardized variance decomposition matrix $\hat{\mathbf H}^{t} = [\hat{h}_{ij}^{t}]$ as the shares of the $K$-step-ahead error variances in forecasting $\mathbf r_{i}$ due to the shocks to $\mathbf r_{j}$,
\begin{equation}\label{GVD}
\begin{aligned}
\hat{h}_{ij}^{t} & =  \frac{\nu_{ij}^{t}}{ \displaystyle\sum_{j = 1}^n \nu_{ij}^{t}}
\quad \text{with} \quad
\nu_{ij}^{t} & = \frac{\sigma_{jj}^{-1}  \displaystyle\sum_{k = 0}^{K-1} \left( \mathbf e_i^{'} \boldsymbol \Theta_k \boldsymbol \Sigma \mathbf e_j \right)^2 }{   \displaystyle\sum_{k = 0}^{K-1} \left( \mathbf e_i^{'} \boldsymbol \Theta_k \boldsymbol \Sigma \boldsymbol \Theta_k^{'} \mathbf e_i \right) }, \\
\end{aligned}
\end{equation}
where $\mathbf e_j$ is a $n$-dimensional vector with the $j$th element unity and zeros elsewhere and $\sigma_{jj}$ is the $j$-th diagonal element of $\boldsymbol \Sigma$. 

To convert this variance decomposition matrix or the similarity matrix to a dissimilarity matrix $\mathbf H^{t, GVDD}=[h_{ij}^{t, GVDD}]$ we use the following formula:
\[
h_{ij}^{t, GVDD} = \sqrt{2(1- \hat{h}_{ij}^t)}.
\]
In this case if two companies are more related to each other, then $h_{ij}^{t,GVDD}$ is smaller.

\section{Graphs}

In this section, we consider the network of financial returns using the graph theory  \cite{bondy}.
A graph or a network $G(V, E)$ contains a set of vertices ($V$) and the relation between the nodes which is indicated by $E$ (see, e.g., \cite{prof4, bondy}). In graphs sometimes there are more than one edge between some nodes, and also there maybe a loop, i.e., an edge from a node to itself. If there is no loop nor multiple edges in the graph we have a \emph{simple graph}. If the relation is symmetric we have an \emph{undirected graph}, otherwise we have a \emph{directed graph}. If the edges have some numeric values (or weights) we have a \emph{weighted graph}. Based on the measures proposed in Section \ref{Sec:dissimilarity}, the smaller weights (or lengths) of the edges means that they are more similar. For each graph we have some subgraphs as well. A sub-graph of a graph $G(V,E)$ is a graph $G'=(V',E')$ such that $V'\subset V$ and $E'\subset E$. Another characteristic of a graph is the conceitedness. We say that an undirected graph $G$ is \emph{connected}, if and only if there is a path between all the nodes in the graph. A similar definition is also presented for a directed graph which is called \emph{strongly connected graph}. For further definitions and properties from graph theory we refer to \cite{prof4}.

 \subsection{Adjacency Matrix}

\begin{figure}[tbph]
\begin{center}
\includegraphics[height=5cm]{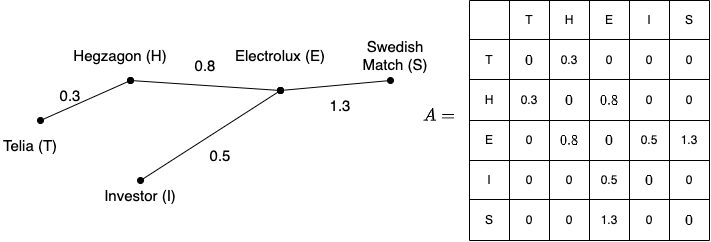}
\end{center}
\vspace*{-0.2in}\caption{Undirected graph consisting of five vertices together with the adjacency matrix $\mathbf A$. In the case of the PCCD, the elements $A_{ij}$ of the adjacency matrix $\mathbf A$ correspond to the artificial elements of $\mathbf H^{t,PCCD}$ for some $t$ which are random numbers between 0 and 2. 
\label{fig:3}}
\end{figure}

One way to describe a graph is by an adjacency matrix. For a given graph with $n$ vertices $V_1, V_2, ..., V_n$, the adjacency matrix $A$ is an $n$ by $n$ matrix such that the numbers of row and columns is equal to the number of vertices $V$. Then, the element $(i,j)$ in the adjacency matrix is $w_{i,j}$ (or $0$) if there is (or is not) an edge between vertices $V_i$ and $V_j$ with the weight of $w_{i,j}$. The adjacency matrix is symmetric if the graph is undirected  which means that $A_{i,j} = A_{j,i}$. In Figure \ref{fig:3}, an illustration of a graph is presented, which is constructed by using the artificial returns on five stocks traded on the Swedish capital market.

Contrariwise, for any squared matrix $A$ we can construct a graph that the number of vertices of the graph is equal to the number of columns of the matrix and between any two vertices $V_i$ and $V_j$ we add an edge with the weight of $A_{
i,j}$. This matrix $A$ is the adjacency matrix of the graph. In the following, we consider two measures to find the relationship between companies: PCCD and GVDD.
The adjacency matrix that we obtain from the PCCD is a symmetric matrix and we construct an undirected graph. Similarly, we construct an asymmetric adjacency matrix and a directed graph by using the GVDD measure. 

\subsection{Distance Matrix}
\begin{figure}[tbph]
\begin{center}
\includegraphics[height=5cm]{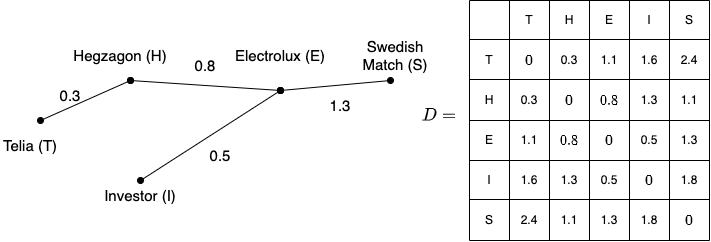}
\end{center}
\vspace*{-0.2in}\caption{
Undirected graph consisting of five vertices together with the distance matrix $\mathbf D$.
\label{fig:6}}
\end{figure}

For a graph we can construct the distance matrix by using adjacency matrix.
The distance between two vertices $V_i$ and $V_j$ is the summation of weights of all the edges in the shortest path from $V_i$ to $V_j$. The distance matrix $D$ for a graph $G$ is a squared matrix such that the number of columns is the number of vertices of the graph and each element $D_{i,j}$ indicates the distance from vertex $V_i$ to $V_j$. If the graph is undirected, the distance matrix is a symmetric matrix, otherwise the matrix is asymmetric. Figure \ref{fig:6} depicts the undirected graph and the distance matrix obtained by using the results presented in Figure \ref{fig:3}.

In this section, we are interested in graphs and the clustering of the networks that we make by using the variance decomposition matrix. Networks and graphs are two important topics in the field of statistics and finance and have attracted lots of attentions (see, \cite{Mantegna1999,Diebold2014, GTRoEIaTCM, RCC}).

\subsection{Center of a Graph}

There are some characteristics of graphs that we use in this paper. One of the important property of the graph is the center of the graph which we define below (see also \cite{center} for more information and definition related to the center of the graph).


In a network of companies, we consider that the center of the network is a vertex (or a set of vertices) in the graph that has a minimum value of the maximum distances from it (them) to other vertices. For finding the center of a graph, we add a column (or a row) to the distance matrix of the graph, called \emph{max}, whose elements $\text{max}_i$ indicate the maximum distance from the $i$-th vertex to other vertices. The center of the graph is then the vertex (a set of vertices) that has the minimum value at the \emph{max} column.

\begin{figure}[tbph]
\begin{center}
\includegraphics[height=5cm]{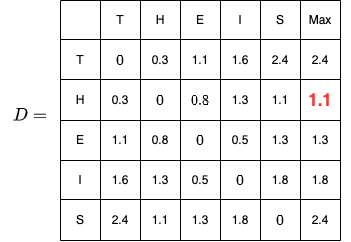}
\end{center}
\vspace*{-0.2in}\caption{Determination the center of the graph in Figure \ref{fig:6}.  The added column to the distance matrix $\mathbf D$ is \emph{max}. The minimum value of this column, depicted red, specifies \emph{Hegzagon} as the center of the graph.
\label{fig:7}}
\end{figure}

Figure \ref{fig:7} illustrates the computation of the center of the undirected graph of Figure \ref{fig:6}, which appears to be \emph{Hegzagon}.

Based on the results discussed in the previous section, it can be proven that the PCCD is a distance. Therefore, at each window $t$ and for any pair of $i$ and $j$, ${h}^{t,PCCD}_{i,j}$ is the shortest distance between two vertices $V_i$ and $V_j$. Therefore, for the PCCD method, the distance matrix is equal to the adjacency matrix.

\subsection{Hierarchical Clustering of a Data}

As the graph of the financial assets is a presentation of dense network structure, analysing and reporting their properties based on different connectedness measures can be very difficult. Instead, we work with a tree clustering that indicates the relation between companies.
There are many methods to cluster the data, for example, flat clustering and hierarchical clustering (see, \cite{ACFfSBHC, AICFfHCT}). We focus on a hierarchical clustering in this paper because of some advantages of the method. The first advantage of using hierarchical clustering is that we do not need to indicate the numbers of clusters before starting the clustering. Another advantage is that the structure of the cluster is a tree and therefore we can use some properties of trees, such as distance between trees. 
In the hierarchical clustering tree, the leaves correspond to the firms and each internal node corresponds to a cluster such that all the data in one cluster are indicated by the leaves of the subtree rooted at the internal node. 

There are two methods for hierarchical clustering: \emph{agglomerative} and \emph{divisive} \cite{ACFfSBHC, AICFfHCT}. The algorithm for the divisive method is more complicated than the one for the agglomerative method. Moreover, most of them are NP-hard to compute which means that there is no known polynomial time algorithm for implementing them. As such, we make use of agglomerative methods. We first start by vertices that are more similar to each other and then merge them until we reach to the groups that are less similar to each other. At last, we merge even those groups that are completely different from each other. In all kinds of hierarchical clustering all the groups are merged eventually.

In this paper choose to work with single linkage clustering algorithm which is efficient and suitable for symmetric distance. 
When the matrix is asymmetric, then we consider the max between the element $uv$ and $vu$ in the asymmetric matrix and we convert it to a symmetric matrix. Then, the methods that exist for symmetric matrices, are employed (see, \cite{HCoAN} for details).



\subsection{Distance between Trees}
\label{distance}
In the previous section, different types of clustering methods were introduced for the stock returns. As in each period, a hierarchical clustering tree is obtained and the changes in the hierarchical clustering tree can be summarized by using the distances between the trees. We start this section by introducing some methods for calculating the distance between trees.

The \emph{tree edit distance} and the \emph{tree alignment distance} are two distances that are primarily defined between trees (see, \cite{ASoTEDaRP}). Furthermore, the \emph{interleaving distance} and the \emph{Frechet-like distance} are defined between \emph{merge trees}\footnote{From \cite{IDbMT} a merge tree is a rooted tree with a real valued function which is defined on the tree. The function is monotonically decreasing from the root the leaves. } (see, e.g., \cite{IDbMT, FLDbTRT}). Recently, the interleaving distance were generalized in \cite{FAfCGHaIDbT} who proposed a fixed parameter tractable algorithm for finding the interleaving distance between two merge trees. The generalized Robinson-Foulds metrics for comparing and finding the similarity between \emph{phylogenetic trees}\footnote{ A phylogenetic tree is a rooted labeled tree such that the tree indicates the evolutionary relation between different species.} has been worked by M.R. Smith. The practical computation of the distance between trees can be performed by using the R package \emph{TreeDist} (see, \cite{ITGRFMfCPT}).




\subsubsection{Robinson-Foulds Distance} 
\emph{Robinson-Foulds} distance is a distance which is defined on unrooted labeled trees.
Each edge in a
tree is a bridge \footnote{In a graph a bridge is an edge that if we cut it, the draph is divided into two separate graphs} that divides the leaves of a labeled tree into two groups such that there is no overlapping between them. The Robinson-Foulds algorithm counts the number of splits in one tree that does not exist in another one (see, \cite{MSDfUBPT, ITGRFMfCPT}). Or in another word, it is defined by

\begin{equation*}
    d_{RF}(T_1, T_2) = \frac{1}{2}|\psi(T_1) \ominus \psi(T_2)|
\end{equation*}
such that $\psi(T_1)$ is the set of all splits related to edges of $T_1$, similar for $\psi(T_2)$. Also, for two sets $A$ and $B$, $A\ominus B = (A\setminus B)\bigcup (B\setminus A)$.

Since the set of rooted trees is a subset of trees, we can also use the above definition for labeled rooted trees. Moreover, the hierarchical clustering trees are labeled rooted trees. As such, we can find the dissimilarity between them or in another word the distance between them by using the Robinson-Foulds distance.

\begin{figure}[htbp]
\begin{center}
\includegraphics[width=10cm]{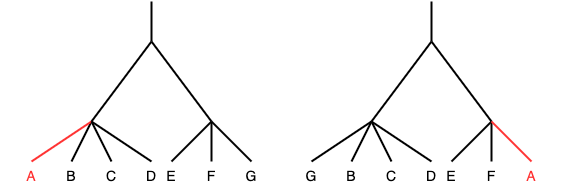}
\end{center}
\vspace*{-0.25in}
\caption{Two trees with similar structures (just one leaf has been changed), but with large Robinson-Foulds distance.
\label{Fig:smaldis}}
\end{figure}

As the Robinson-Foulds method does not provide an acceptable result when there is a small change in trees, for example, when the difference between the two trees $T_1$ and $T_2$ is that just one leaf in $T_1$ moves in $T_2$ like Figure \ref{Fig:smaldis}. In this case the Robinson-Foulds distance returns a very large number that indicates that two trees are not similar. Therefore, the \emph{generalized Robinson-Foulds} method was introduced in \cite{ITGRFMfCPT}.

In \cite{ITGRFMfCPT}, M. R. Smith introduced  
three information based distances between the phylogenetic trees. As we are interested in the distance between the rooted trees that indicates
the relationship between the clustered markets,  the \emph{clustering information distance} is the best suitable one that we can use here. The phylogenetic trees and the hierarchical trees are very similar in the structure. In both of them the leaves of the tree save the information about the data. In the phylogenetic trees we have the name of species, while the name of the companies are used in the hierarchical tree that we have constructed from the relationship between these companies. Also, in phylogenetic trees, the nearest common ancestor of similar species is closer to them rather than the different species. In the hierarchical clustering trees we have the similar situation as well. Namely, if two companies are more related, then they are merged faster than the ones that are more different. Therefore, in this work we use the distance which is defined on the phylogenetic trees for finding the distance between the hierarchical clustering.

\subsubsection{Clustering Information Distance}
The definition of the \emph{clustering information distance} can be found in \cite{ ITGRFMfCPT,CCaIBD,ITMfCC}. In a tree, each edge is a bridge and therefore for each edge $e_{i,j}$ in tree $T_1$ there is a \emph{cut}, 
that is a partition of the leaves of $T_1$ into two partitions $A$ and $B$.
We say that a cut is \emph{trivial} if $|A|~\text{or}~ |B| < 2$ where $|A|$ indicates the number of leaves in the group $A$. A pair of $(
C_{i,j}^1, 
C_{i',j'}^2)$ is called a \emph{pairing} if $C_{i,j}^1$ and $C_{i',j'}^2$ are nontrivial cuts from respectively $T_1$ and $T_2$. A \emph{matching (M)} is a set of pairing in which none of the cuts has been occurred more than once. The score of a matching (M) is the summation of the score of pairing in M. An \emph{optimal matching} is a matching that its score is the highest among all the possible matching between the trees.

Using the above definitions, the clustering information distance is defined as follows. Each edge partitions the leaves of a tree into two clusters $A$ and $B$. If $\pi_A = \frac{|A|}{n}$ (where $n$ is is the number of leaves which in the tree) is the probability that a randomly chosen leaf belongs to the cluster $A$, then for a pair of cuts $C_{i,j}^1$ and $C_{i',j'}^2$ the \emph{mutual clustering information} is defined as follows: $I_{CI}(C_{i,j}^1;C_{i',j'}^2)$ is the probability that a vertex $u$ is in cluster $A'$ (or $B'$) from $C_{i',j'}^2$ by knowing that $u$ is in which cluster in $C_{i,j}^1$ and is earned by the following formula
\[
I_{CL}(C_{i,j}^1;C_{i',j'}^2) =
J(A,A') + J(A,B') + J(B,A') + J(B,B')
\]
where $J(A,B) = \pi_{A\cap B}\times\text{log}(\frac{\pi_{A\cap B}}{\pi_{A} \times\pi_B})$.
By subtracting from a maximum value this measure of similarity is converted to a distance which is called \emph{clustering information distance}.

\section{Empirical illustration}

In this section we consider 28 Swedish companies. We analyze the network structure of the asset returns through the PCCD method and the GVDD method. Then, we find the center of the networks. Also, by using the hierarchical clustering and the information distance between rooted trees, we investigate the changes of the hierarchical trees.  

We first take the adjusted closing prices of 28 Swedish companies from Yahoo Finance for five years, from March 31st, 2017 to March 30th, 2022. In the analysis, a moving window of three months is employed. Commonly, three months have 63 open days and, therefore, we consider the first 63 days as the first window. Then we shift by one day and the second window starts from day two and ends at day 64, and so on. For each window we use two methods to find the adjacency matrices for these companies: (i) PCCD which constructs a symmetric matrix and therefore undirected graph, and (ii) GVDD which constructs an asymmetric matrix and a directed graph, respectively. 

\subsection{Networks of financial returns}

\begin{figure}[tbph]
\begin{center}
\includegraphics[height=8cm]{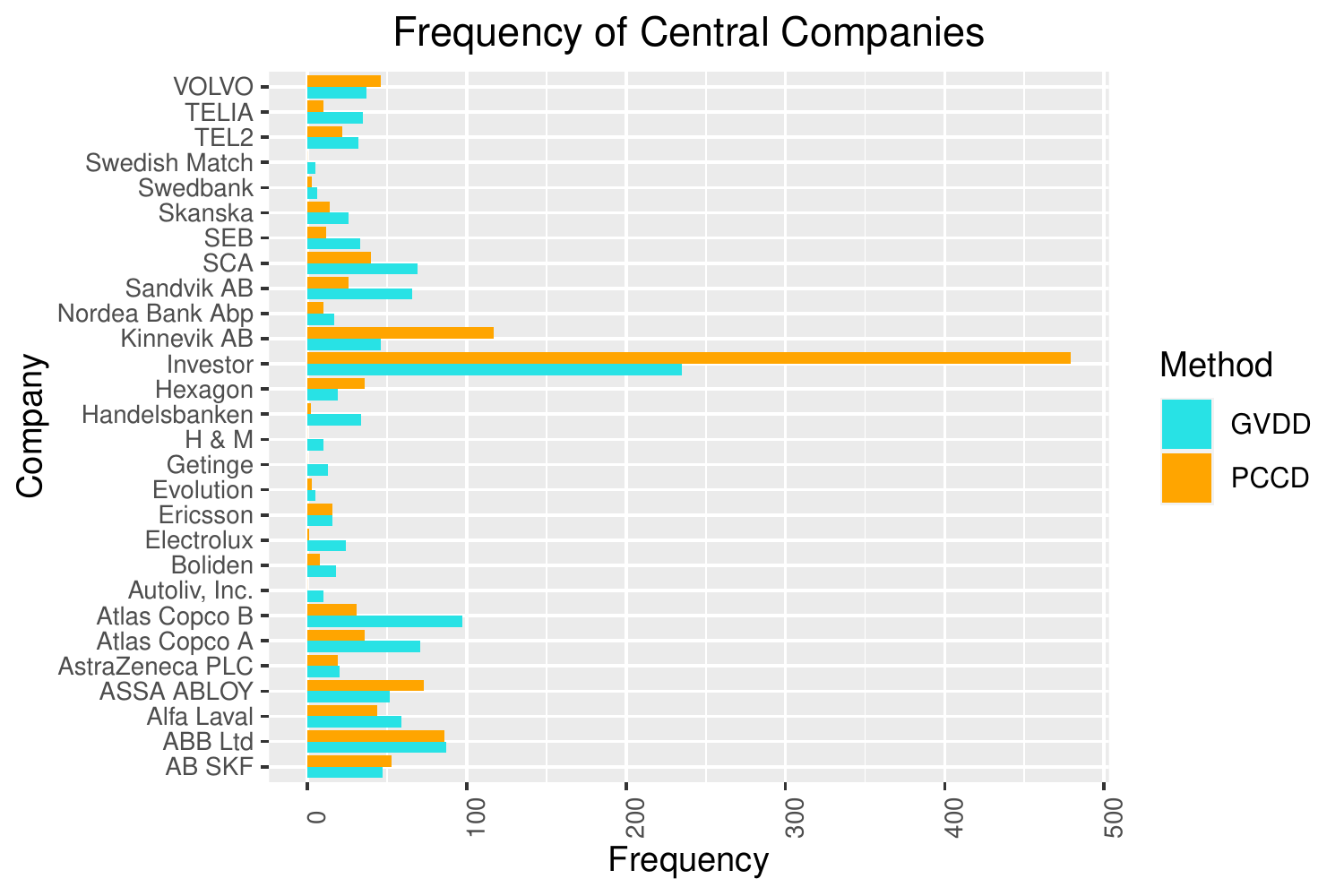}
\end{center}
\vspace*{-0.3in}\caption{
The frequency of the companies to be the center of the graphs using the PCCD method (blue) and the GVDD method (yellow).
\label{fig:8}}
\end{figure}

Using the two proposed methods of determining the adjacency matrix, we find the center of the graphs at each window and compute the frequency of each company to be the center during the past five years. The results are depicted in Figure \ref{fig:8}. 
By definition, the center of the graphs is a company that has most influence to stock returns of all the other companies in the shortest time.

In Figure \ref{fig:8} we can see that using both of the methods the highest frequency of the centers happens for the company which is called \emph{Investor}. It means that during the past five years \emph{Investor} was in most of the time the center of the companies between the 28 companies that we chose. Therefore, as \emph{Investor} is an investment company, the financial industry has the most influence on all the other companies in shortest time in Sweden. We also note that \emph{Sandvik AB} appears to be the company with the second highest frequency by using both the methods.

If we consider the data from 31 of January 2020 to 31 of July 2020, which is the time that COVID-19 was started, \emph{Svenska Cellulosa Aktiebolaget (SCA)} was the most popular center by using the PCCD method and \emph{Investor} was the most popular center by the GVDD method. Also, from 1 October 2020 to 10 of May 2021, which corresponds to the time when the Coronavirus Delta variant was dominant, \emph{Investor} was indicated to be the most popular center by both the methods. Finally, from 1 December 2021 until the last day in the data ASSA ABLOY AB was the most popular center by using both the methods.

\subsection{Hierarchical clustering tree}

\begin{figure}[tbph]
\begin{center}
\begin{tabular}{ccccc}
\hspace{-1cm}\includegraphics[height=6cm]{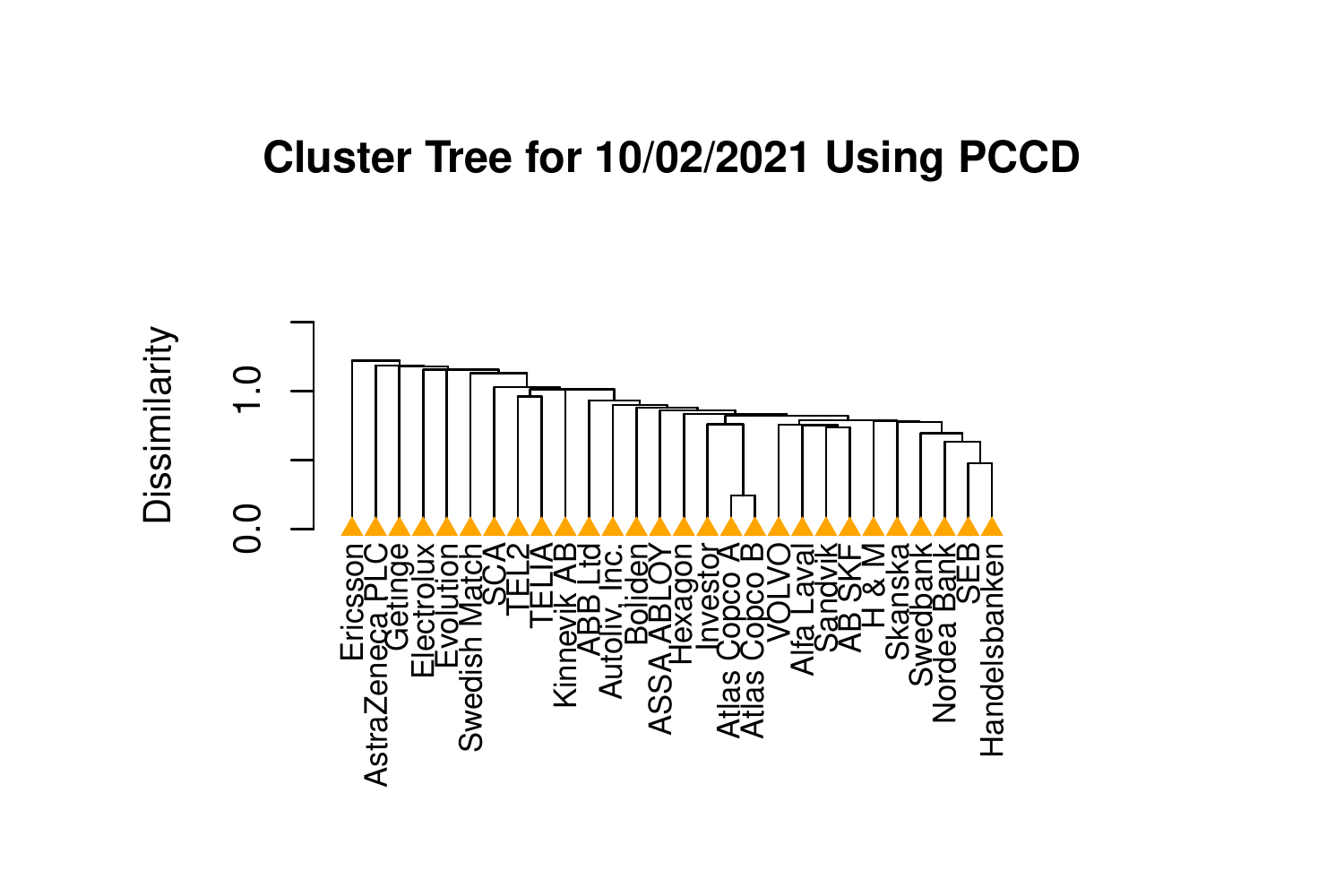} & &
 \hspace{-2cm}\includegraphics[height=6cm]{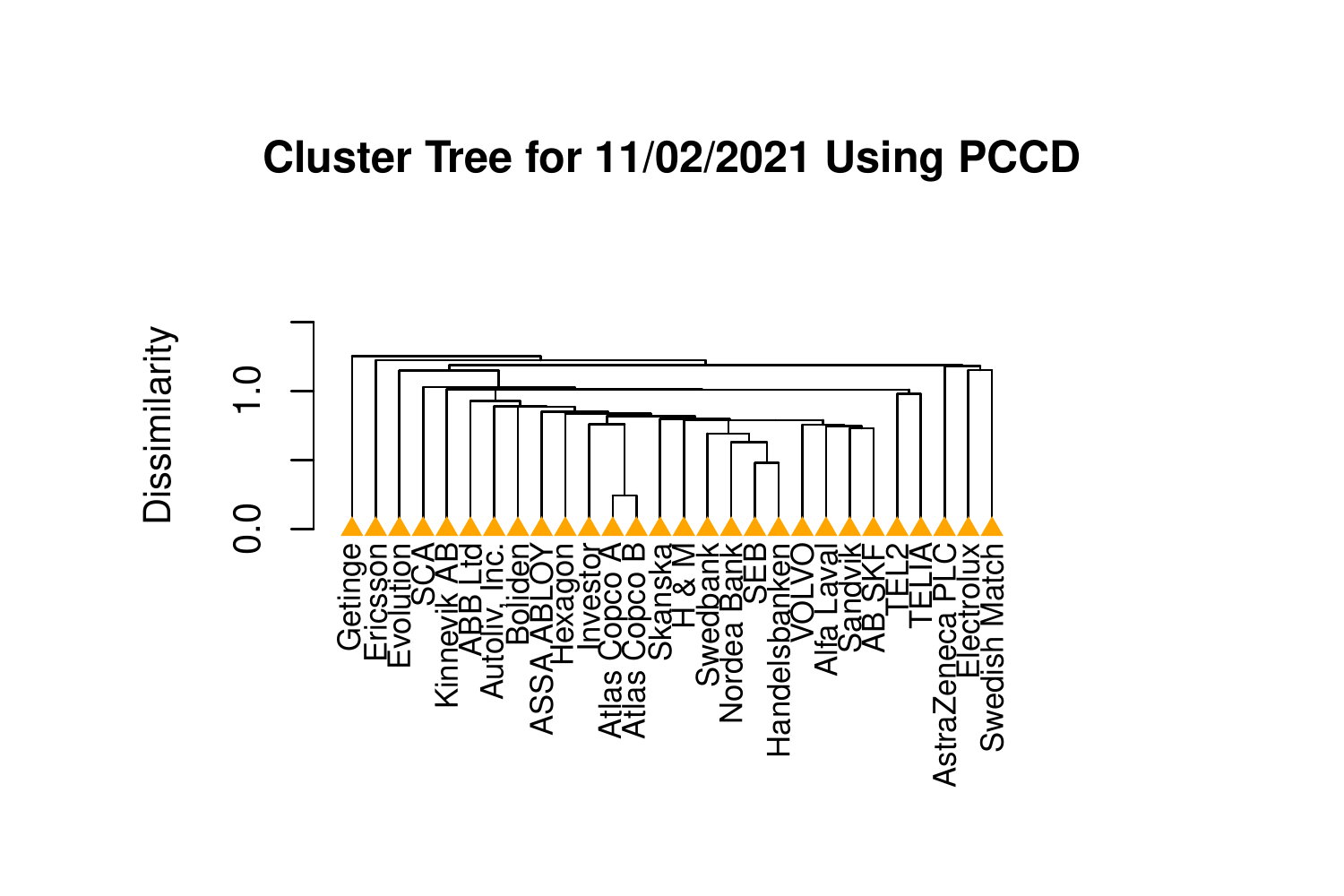}\\
(a) & & (b)
\end{tabular}
\end{center}
\vspace*{-0.13in}\caption{
(a) The hierarchical clustering tree for window number $900$ which is November $8^\text{th}$, $2020$ when the PCCD method is used. (b) The hierarchical clustering tree for the window number $901$ by using the PCCD method. During the last five years, these days had the most different hierarchical trees by using the PCCD method.
\label{fig:ct-pccd}}
\end{figure}

\begin{figure}[tbph]
\begin{center}
\begin{tabular}{ccccc}
\hspace{-1cm}\includegraphics[height=6cm]{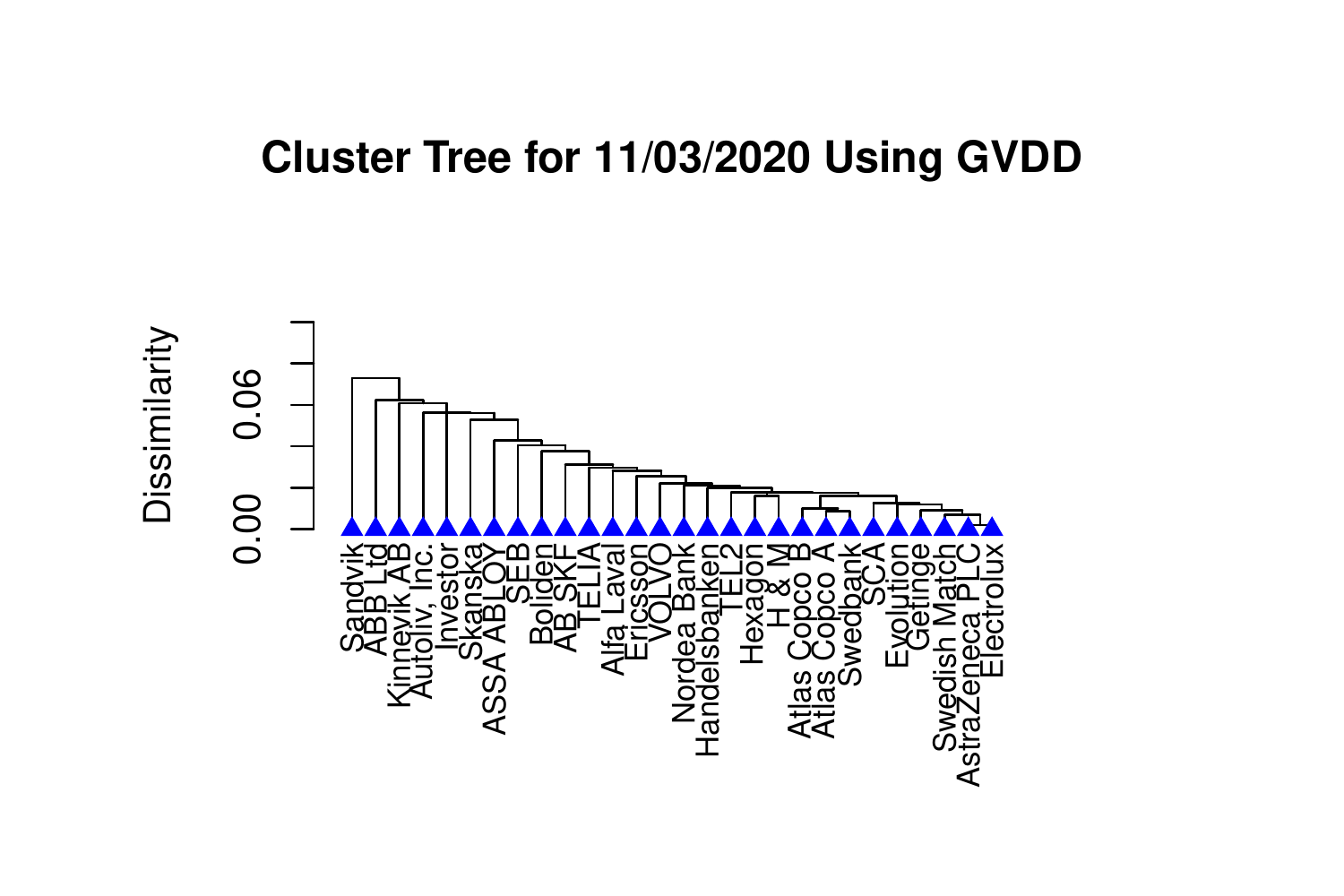} & &
 \hspace{-2cm}\includegraphics[height=6cm]{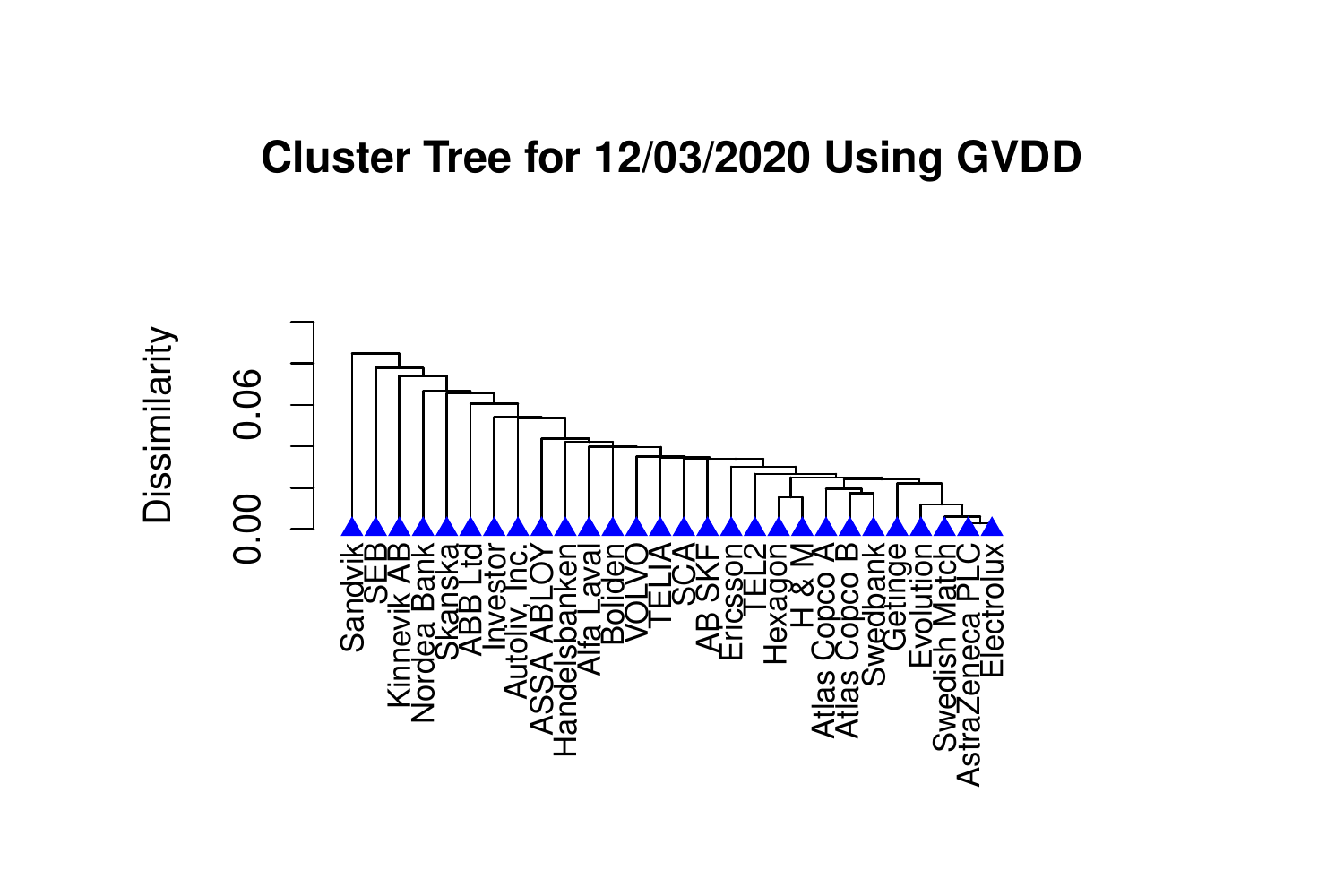}\\
(a) & & (b)
\end{tabular}
\end{center}
\vspace*{-0.13in}\caption{
(a) The hierarchical clustering tree for window number $670$ which is March $11^\text{th}$, $2020$ when the GVDD method is used. (b) The hierarchical Clustering Tree for the window number $671$ by using the GVDD method. During the last five years, these days had the most different hierarchical trees by using the GVDD method.
\label{fig:ct-gvdd}}
\end{figure}

Figures \ref{fig:ct-pccd} and \ref{fig:ct-gvdd} depict the hierarchical clustering trees computed for subsequent days by the PCCD method and the GVDD method. We choose the date that observed the largest changes in the information distance between two hierarchical trees. It took place on 10/02/2021 in the case of the PCCD method, and on 11/03/2020 in the case of the GVDD method.

By looking at these figures we see that there is a big difference between these two methods and their maximum d. First, the height of the hierarchical tistance happens in two different days. Also, we can see that in average the hight of the hierarchical trees in the PCCD method is larger than the one in the GVDD method.

\subsection{Distance between Hierarchical Clustering Trees}
\begin{figure}[htbp]
\begin{center}
\includegraphics[width=12cm]{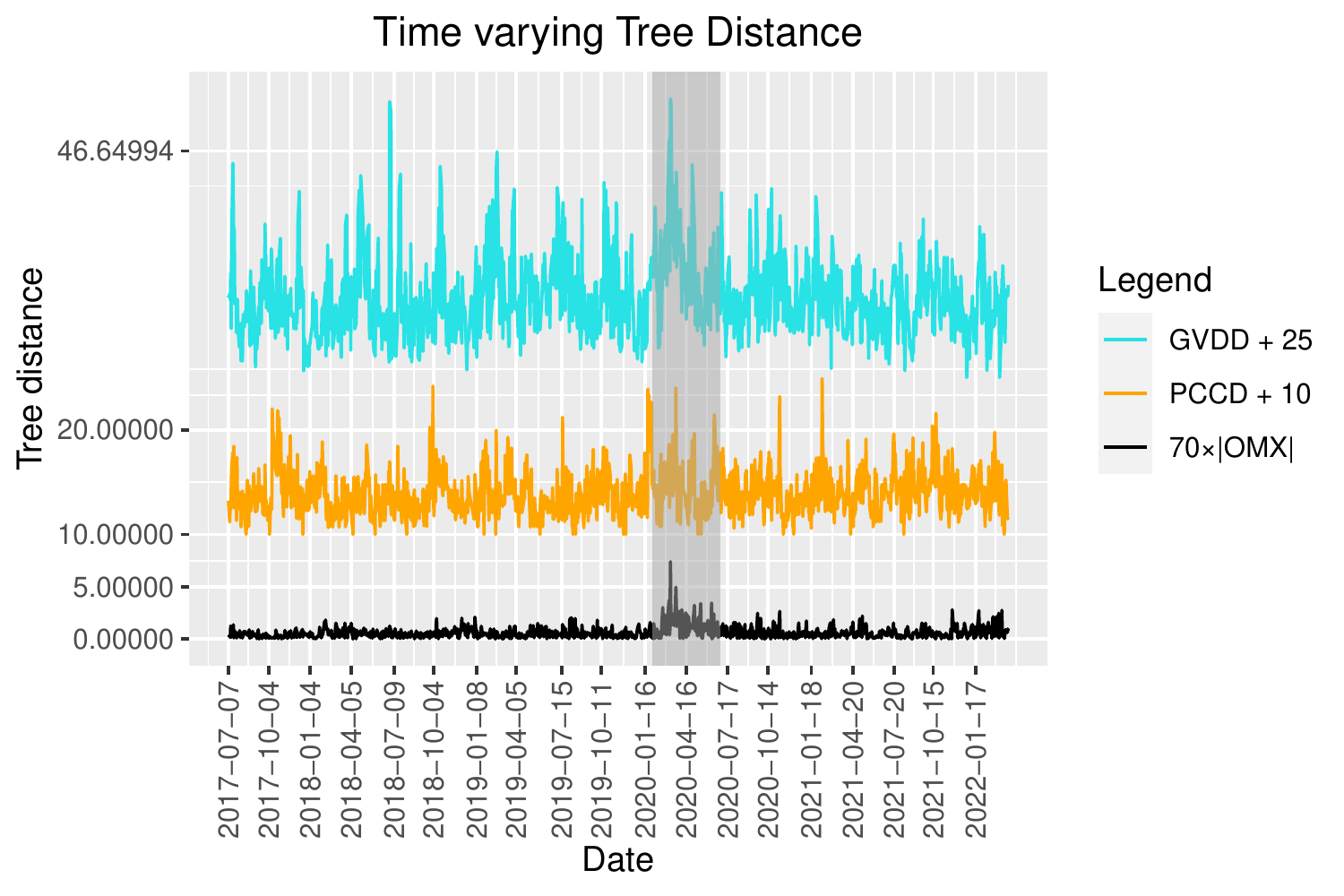}
\end{center}
\vspace*{-0.25in}
\caption{Distances between the hierarchical clustering trees in the GVDD method are depicted blue (the upper one), while the yellow line (the middle one) corresponds to the distances between the hierarchical clustering trees when the PCCD method is used. The dark grey area indicates the time of the first COVID-19 wave. 
\label{fig:clusteringdis}}
\end{figure}

In the previous section, by using the single linkage hierarchical clustering for the PCCD method and for the GVDD method we construct the hierarchical clustering of the stock data at each window. Using the distance defined in Section \ref{distance}, we compute the distance between the hierarchical trees in this section sequentially. The results are depicted in Figure \ref{fig:clusteringdis}.

The three lines in Figure \ref{fig:clusteringdis} demonstrate the behaviour of an autoregressive process. To study this effect and also to investigate possible (lag) relationships between the three time series we fit a vector autoregressive model (VAR) to these series. First, the order of the autoregressive model is chosen by using the Hannan and Quinn model selection criteria (see, \cite{hannan1979determination}), which results in two. Second, we fit a VAR(2) model to $(\{PCCD_t\},\{GVDD_t\},\{OMX_t\})$, which leads to the following multivariate model

\begin{eqnarray*}
\begin{pmatrix}
PCCD_t\\GVDD_t\\OMX_t\\
\end{pmatrix}
&=&
\begin{pmatrix}
1.72767^{***}\\5.03862^{***}\\0.00146\\
\end{pmatrix}
+
\begin{pmatrix}
0.20913^{***}&0.04571^{*}&-10.91973^{*}\\
0.19580^{***}&0.31478^{***}&-25.64787^{**}\\
0.00009&-0.00011&-0.04697\\
\end{pmatrix}
\begin{pmatrix}
PCCD_{t-1}\\GVDD_{t-1}\\OMX_{t-1}\\
\end{pmatrix}\\
&+&
\begin{pmatrix}
0.22751^{***}&0.01034&3.84118\\
0.14021^{**} &0.00758&11.12222\\
0.00014&-0.00011&-0.01427\\
\end{pmatrix}
\begin{pmatrix}
PCCD_{t-2}\\GVDD_{t-2}\\OMX_{t-2}\\
\end{pmatrix}
+\boldsymbol \epsilon_t,
\end{eqnarray*}
where $\{\boldsymbol \epsilon_t\}$ is a white noice process with covariance matrix given by
\[
\mathbf \Sigma=   \begin{pmatrix}
 4.26874 & 1.21287&-0.00024\\
 1.21287 &11.87265&-0.00321\\
-0.00024 &-0.00321& 0.00014\\
 \end{pmatrix}.
\]

In the model equation, the coefficients denoted with '$^{***}$' are statistically significant at 0.1\%, '$^{**}$' -- at 1\%, '$^{*}$' -- at 5\%, and '$^{.}$' -- at 10\%. We observe that the current values of tree distances constructed by using the PCCD and GVDD methods are positively correlated with their previous values. Moreover, the values obtained by using the PCCD method have also impact on future values obtained for both the PCCD method and the GVDD method at lag 2. While the previous distances are positively correlated with the future ones, the values of the Swedish capital market index, OMX, have negative significant impact at lag 1. To this end, we note the OMX index cannot be predicted by none of the distances considered in the study neither by the previous values of the index itself.

\begin{figure}[tbph]
\begin{center}
\begin{tabular}{ccccc}
\hspace{-1cm}\includegraphics[height=5.5cm]{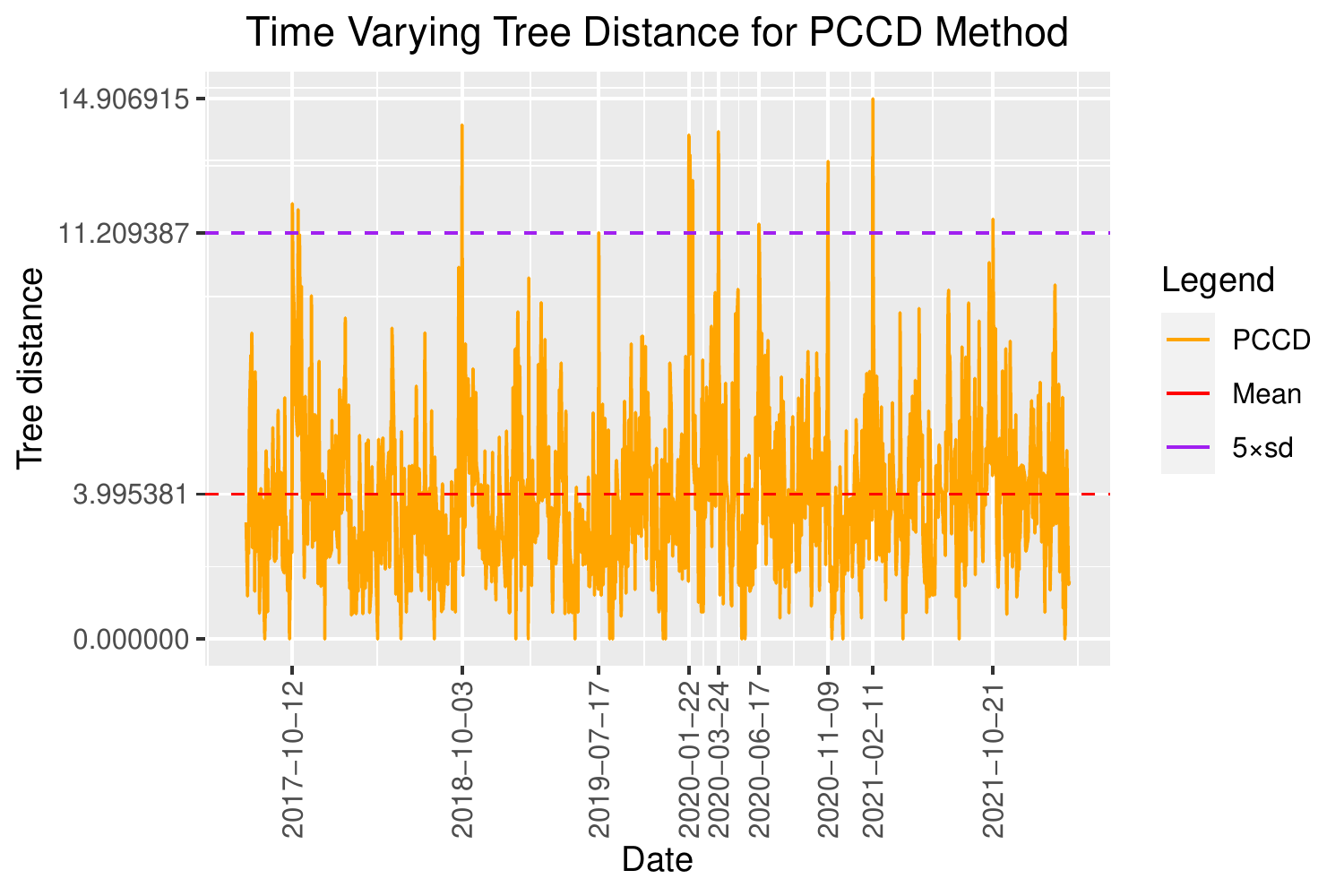} & ~~~~~~~~~~~ &
 \hspace{-2cm}\includegraphics[height=5.5cm]{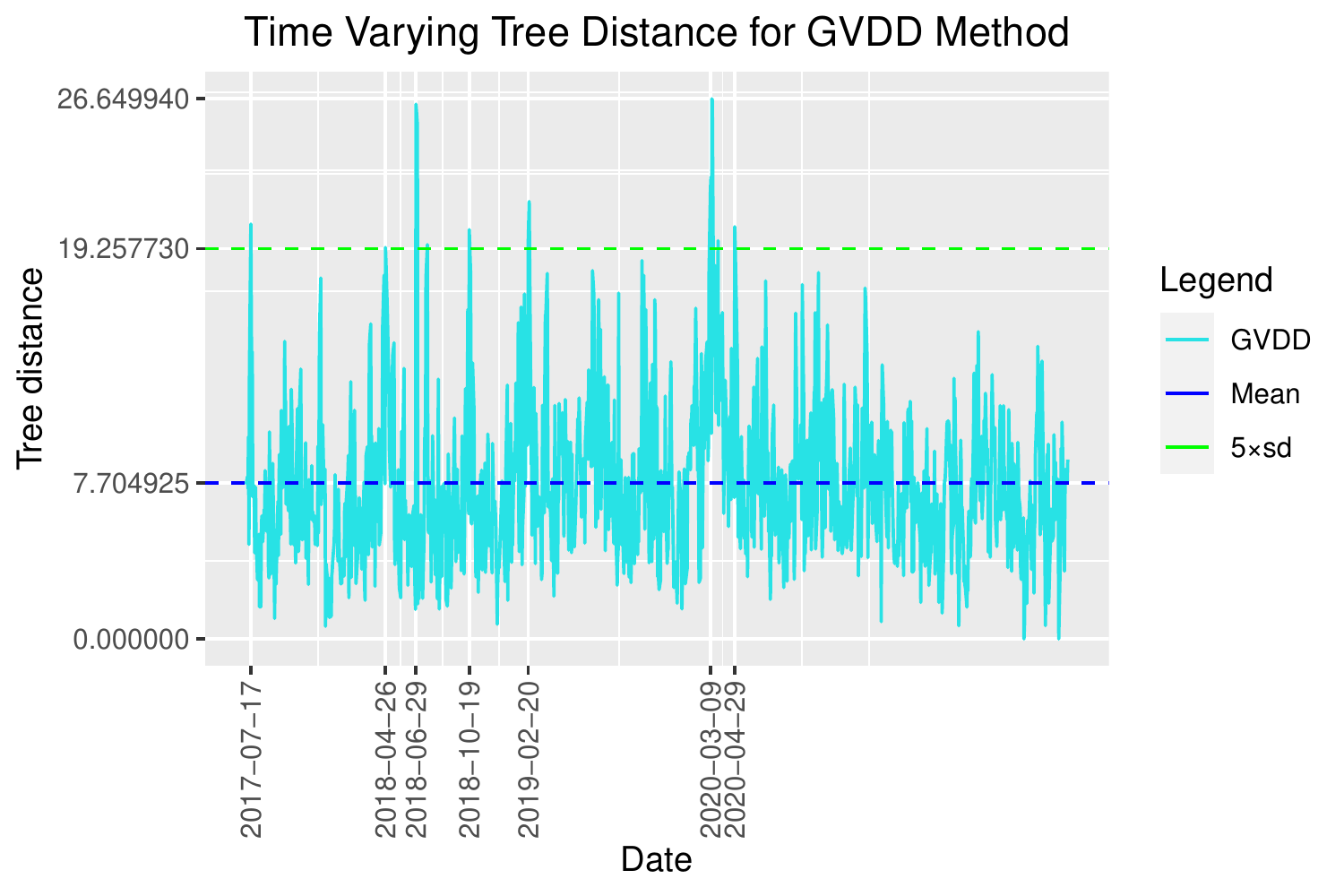}\\
(a) & & (b)
\end{tabular}
\end{center}
\vspace*{-0.13in}\caption{
(a) Distances between the hierarchical clustering trees computed by using PCCD method. The purple dashed line indicates $5\times$SD and the red dashed line is the mean line. (b) Distances between the hierarchical clustering trees obtained by using the GVDD method. The green dashed line indicates $5\times$SD and the blue dashed line is the mean line.
\label{fig:2picture}}
\end{figure}

In Figure \ref{fig:clusteringdis}, we compare the methods PCCD and GVDD methods and we see that, in general, the application of the GVDD method leads to larger values of the distances between the trees over time, which means that in the GVDD methods the hierarchical clustering trees differ more than in the case when the hierarchical clustering trees are constructed by using the PCCD method. Also, the average value of the distances is larger when the GVDD method is used in comparison to the PCCD method. Both plots in Figure \ref{fig:2picture} depict the series with the distances together with the mean line and $5 \times$ standard deviation (SD) line. The comparison of the pattern of computed distances with $5 \times$SD corresponds to the application of the Shewhart control chart for detecting changes in statistical process control (see, e.g., \cite{psarakis2007spc,bodnar2011cusum,bisiotis2021control}). We see that most of the times that the jumps of magnitude larger than $5 \times$SD happen after the beginning of the year $2020$ in both plots. Also, the numbers of jumps that are higher than $5\times$SD is larger when the GVDD method is used. Moreover, $7$ out of $20$ jumps that are above the green line happen in 2020 during the COVID-19 time.

\section{Conclusion}

Specifying and monitoring the structure of financial market is an important research topic with direct applications to real-life problem. The knowledge about he center of the capital market is useful in the determination of its stability, while clustering the companies on the capital market provides us information about the connectedness of the companies traded on it. The above mentioned topics are treated in the literature by constructing a graph and determining its center as well as by finding the hierarchical clustering trees.

In the paper, two methods are compared for determining the network of companies traded on the Swedish capital market. While the first approach, the PCCD method, is based on a symmetric adjacency matrix, the second one, the GVDD method, employs an asymmetric adjacency matrix. Both methods indicate the company \emph{Investor} as the center of the Swedish capital market in most of the considered cases. On the other side, the company \emph{Swedish Match} shows the largest dissimilarity to the all other companies traded on the Swedish stock exchange. Finally, computing the distances between the hierarchical clustering trees we found the most of the changes in the structure of the Swedish capital market happen at the beginning of 2020, i.e., during the first COVID-19 wave.

\bibliographystyle{abbrv}
\bibliography{Network-Aug19-2022}

\begin{thebibliography}{10}

\bibitem{ASoTEDaRP}
P.~Bille.
\newblock A survey on tree edit distance and related problems.
\newblock {\em Theoretical Computer Science}, 337(1-3):217--239, June 2005.

\bibitem{bisiotis2021control}
K.~Bisiotis, S.~Psarakis, and A.~N. Yannacopoulos.
\newblock Control charts in financial applications: An overview.
\newblock {\em Quality and Reliability Engineering International}, 2021.

\bibitem{bodnar2011cusum}
O.~Bodnar and W.~Schmid.
\newblock Cusum charts for monitoring the mean of a multivariate gaussian
  process.
\newblock {\em Journal of Statistical Planning and Inference},
  141(6):2055--2070, 2011.

\bibitem{MSDfUBPT}
D.~Bogdanowicz and K.~Giaro.
\newblock Matching split distance for unrooted binary phylogenetic trees.
\newblock {\em IEEE-ACM Transactions on Computational Biology and
  Bioinformatics}, pages 1--12, 2011.

\bibitem{Bonanno2004}
G.~Bonanno, G.~Caldarelli, F.~Lillo, S.~Micciche, N.~Vandewalle, and R.~N.
  Mantegna.
\newblock Networks of equities in financial markets.
\newblock {\em The European Physical Journal B}, 38(2):363--371, 2004.

\bibitem{bondy}
J.~A. Bondy and U.~S.~R. Murty.
\newblock {\em Graph Theory with Applications}.
\newblock Elsevier, New York, 1976.

\bibitem{Bouri2021}
E.~Bouri, O.~Cepni, D.~Gabauer, and R.~Gupta.
\newblock Return connectedness across asset classes around the covid-19
  outbreak.
\newblock {\em International Review of Financial Analysis}, 73:101646, 2021.

\bibitem{HCoAN}
G.~Carlsson, F.~M{\'e}moli, A.~Ribeiro, and S.~Segarra.
\newblock Hierarchical clustering of asymmetric networks.
\newblock {\em Advances in Data Analysis and Classification}, 12(1):65--105,
  2018.

\bibitem{GTRoEIaTCM}
L.~F.~D. Cerbo and S.~Taylor.
\newblock Graph theoretical representations of equity indices and their
  centrality measures.
\newblock {\em Quantitative Finance}, 21(4):523--537, 2021.

\bibitem{Chi2010}
K.~T. Chi, J.~Liu, and F.~C. Lau.
\newblock A network perspective of the stock market.
\newblock {\em Journal of Empirical Finance}, 17(4):659--667, 2010.

\bibitem{prof4}
T.~H. Cormen, C.~E. Leiserson, R.~L. Rivest, and C.~Stein.
\newblock {\em Introduction to Algorithms}.
\newblock The MIT Press, 2nd edition, 2001.

\bibitem{ACFfSBHC}
S.~Dasgupta.
\newblock A cost function for similarity-based hierarchical clustering, 2015.

\bibitem{Diebold2014}
F.~X. Diebold and K.~Y{\i}lmaz.
\newblock On the network topology of variance decompositions: Measuring the
  connectedness of financial firms.
\newblock {\em Journal of Econometrics}, 182(1):119--134, 2014.

\bibitem{Diebold2015}
F.~X. Diebold and K.~Y{\i}lmaz.
\newblock {\em Financial and macroeconomic connectedness: A network approach to
  measurement and monitoring}.
\newblock Oxford University Press, USA, 2015.

\bibitem{FLDbTRT}
E.~Farahbakhsh~Touli.
\newblock Fr\'echet-like distances between two rooted trees.
\newblock {\em Journal of Algorithms and Computation}, 53(1):1--12, 2021.

\bibitem{hannan1979determination}
E.~J. Hannan and B.~G. Quinn.
\newblock The determination of the order of an autoregression.
\newblock {\em Journal of the Royal Statistical Society: Series B
  (Methodological)}, 41(2):190--195, 1979.

\bibitem{Koop1996}
G.~Koop, M.~H. Pesaran, and S.~M. Potter.
\newblock Impulse response analysis in nonlinear multivariate models.
\newblock {\em Journal of Econometrics}, 74(1):119--147, 1996.

\bibitem{Mantegna1999}
R.~N. Mantegna.
\newblock Hierarchical structure in financial markets.
\newblock {\em The European Physical Journal B-Condensed Matter and Complex
  Systems}, 11(1):193--197, 1999.

\bibitem{CCaIBD}
M.~Meilă.
\newblock Comparing clusterings—an information based distance.
\newblock {\em Journal of Multivariate Analysis}, 98(5):873--895, 2007.

\bibitem{Minoiu2015}
C.~Minoiu, C.~Kang, V.~Subrahmanian, and A.~Berea.
\newblock Does financial connectedness predict crises?
\newblock {\em Quantitative Finance}, 15(4):607--624, 2015.

\bibitem{IDbMT}
D.~Morozov, K.~Beketayev, and G.~H. Weber.
\newblock Interleaving distance between merge trees.
\newblock In {\em Workshop on Topological Methods in Data Analysis and
  Visualization: Theory, Algorithms and Applications}, 2013.

\bibitem{Nye2006}
T.~M. Nye, P.~Lio, and W.~R. Gilks.
\newblock A novel algorithm and web-based tool for comparing two alternative
  phylogenetic trees.
\newblock {\em Bioinformatics}, 22(1):117--119, 2006.

\bibitem{Onnela2003}
J.-P. Onnela, A.~Chakraborti, K.~Kaski, J.~Kertesz, and A.~Kanto.
\newblock Asset trees and asset graphs in financial markets.
\newblock {\em Physica Scripta}, 2003(T106):48, 2003.

\bibitem{Onnela2004}
J.-P. Onnela, K.~Kaski, and J.~Kert{\'e}sz.
\newblock Clustering and information in correlation based financial networks.
\newblock {\em The European Physical Journal B}, 38(2):353--362, 2004.

\bibitem{Pesaran1998}
H.~H. Pesaran and Y.~Shin.
\newblock Generalized impulse response analysis in linear multivariate models.
\newblock {\em Economics Letters}, 58(1):17--29, 1998.

\bibitem{psarakis2007spc}
S.~Psarakis and G.~Papaleonida.
\newblock Spc procedures for monitoring autocorrelated processes.
\newblock {\em Quality Technology \& Quantitative Management}, 4(4):501--540,
  2007.

\bibitem{ITGRFMfCPT}
M.~R. Smith.
\newblock {Information theoretic generalized Robinson–Foulds metrics for
  comparing phylogenetic trees}.
\newblock {\em Bioinformatics}, 36(20):5007--5013, 07 2020.

\bibitem{RCC}
E.~F. Touli and O.~Lindberg.
\newblock Relative clustering coefficient.
\newblock {\em Journal of Algorithms and Computation}, 54:99--108, 2022.

\bibitem{FAfCGHaIDbT}
E.~F. Touli and Y.~Wang.
\newblock {FPT}-algorithms for computing gromov-hausdorff and interleaving
  distances between trees.
\newblock {\em Journal of Computational Geometry}, 13:89--124, 2022.

\bibitem{Vandewalle2001}
N.~Vandewalle, F.~Brisbois, and X.~Tordoir.
\newblock Non-random topology of stock markets.
\newblock {\em Quantitative Finance}, 1(3):372, 2001.

\bibitem{ITMfCC}
N.~X. Vinh, J.~Epps, and J.~Bailey.
\newblock Information theoretic measures for clusterings comparison: Variants,
  properties, normalization and correction for chance.
\newblock {\em Journal of Machine Learning Research}, 11:2837–2854, Dec.
  2010.

\bibitem{AICFfHCT}
D.~Wang and Y.~Wang.
\newblock An improved cost function for hierarchical cluster trees.
\newblock {\em Journal of Computational Geometry}, 11(1):283--331, 2020.

\bibitem{center}
S.~Wasserman and K.~Faust.
\newblock {\em Social network analysis: Methods and applications}, volume~8.
\newblock Cambridge university press, 1994.

\end{thebibliography}

\end{document}